\documentclass[usenatbib]{mn2e}

\usepackage{amsmath}
\usepackage{graphicx}
\usepackage{deluxetable}
\usepackage{savesym}
\usepackage{amsmath}
\savesymbol{iint}
\usepackage{txfonts}

\usepackage{subfigure}

\def\simless{\mathbin{\lower 3pt\hbox
{$\rlap{\raise 5pt\hbox{$\char'074$}}\mathchar"7218$}}}   
\def\simmore{\mathbin{\lower 3pt\hbox
{$\rlap{\raise 5pt\hbox{$\char'076$}}\mathchar"7218$}}}   
\newcommand{\be}{\begin{equation}}
\newcommand{\ee}{\end{equation}}
\newcommand       \bea          {\begin{eqnarray}}
\newcommand       \eea          {\end{eqnarray}}
\newcommand       \apj          {ApJ}
\newcommand       \apjl         {ApJL}
\newcommand       \aap          {A\&A}
\newcommand       \nat          {Nature}
\newcommand       \mnras        {MNRAS}

\newcommand       \aj      {AJ}

\newcommand       \icarus      {Icarus}
\newcommand       \araa      {ARA\&A}
\newcommand       \nar   {New Astronomy Reviews}

\newcommand      \apjs {ApJ Supplements}

\newcommand \planss{PLANSS}

\newcommand \jgr{JGR}
\newcommand \grl{GRL}

\def\simlt{\mathrel{\hbox{\rlap{\hbox{\lower4pt\hbox{$\sim$}}}\hbox{$<$}}}}
\def\simgt{\mathrel{\hbox{\rlap{\hbox{\lower4pt\hbox{$\sim$}}}\hbox{$>$}}}}
\def\lesssim{\mathrel{\hbox{\rlap{\hbox{\lower4pt\hbox{$\sim$}}}\hbox{$<$}}}}

\def\gtrsim{\mathrel{\hbox{\rlap{\hbox{\lower4pt\hbox{$\sim$}}}\hbox{$>$}}}}

\def\tabby{\rm KIC 8462852}

\title[Dimming of KIC 8462852 following its consumption of a planet]{Secular dimming of KIC 8462852 following its consumption of a planet}

\author[Metzger, Shen \& Stone]{Brian D.~Metzger$^{1}$, Ken J.~Shen$^{2}$, Nicholas Stone$^{1}$ \\
\\$^{1}$Columbia Astrophysics Laboratory, Columbia University, New York, NY, 10027, USA \\$^{2}$Department of Astronomy and Theoretical Astrophysics Center, 
University of California, Berkeley, CA 94720, USA}

\begin{document}
\date{Received / Accepted}
\pagerange{\pageref{firstpage}--\pageref{lastpage}} \pubyear{2017}

\maketitle

\label{firstpage}

\begin{abstract}
The {\it Kepler}-field star \tabby, an otherwise apparently ordinary F3 main-sequence star, showed several highly unusual dimming events of variable depth and duration.  Adding to the mystery was the discovery that \tabby~faded by 14\% from 1890 to 1989 \citep{Schaefer16}, as well as by another 3\% over the 4 year {\it Kepler} mission \citep{Montet&Simon16}.  Following an initial suggestion by Wright \& Sigurdsson, we propose that the secular dimming behavior is the result of the inspiral of a planetary body or bodies into \tabby, which took place $\sim 10-10^{4}$ years ago (depending on the planet mass).  Gravitational energy released as the body inspirals into the outer layers of the star caused a temporary and unobserved brightening, from which the stellar flux is now returning to the quiescent state.  The transient dimming events could then be due to obscuration by planetary debris from an earlier partial disruption of the same inspiraling bodies, or due to evaporation and out-gassing from a tidally detached moon system.  Alternatively, the dimming events could arise from a large number of comet- or planetesimal-mass bodies placed onto high eccentricity orbits by the same mechanism (e.g.~Lidov-Kozai oscillations due to the outer M-dwarf companion) responsible for driving the more massive planets into \tabby.  The required high occurrence rate of \tabby-like systems which have undergone recent major planet inspiral event(s) is the greatest challenge to the model, placing large lower limits on the mass of planetary systems surrounding F stars and/or requiring an unlikely probability to catch \tabby~in its current state.
\end{abstract} 
  

\section{Introduction}

The {\it Kepler}-field star KIC 8462852, an otherwise apparently ordinary F3 main-sequence star with a {\it Gaia}-measured parallax distance of about 400 pc (\citealt{Hippke&Angerhausen16}), was discovered by the Planet Hunters team to exhibit highly peculiar and unique photometric features  \citep{Boyajian+16}.  Over the 4 year period of the nominal {\it Kepler} mission (\citealt{Borucki+10}), the star underwent several dimming events of variable depth and duration, with reductions in the total stellar flux ranging from approximately $0.5-20$\%.  These events are often asymmetric in shape, displaying an apparent lack of periodicity and repeatability.  Groups of clustered dips were observed, centered around days 800 and 1500 (`D800' and `D1500', respectively), but smaller dips are detected at other phases over the observing period.  

Various hypotheses have been put forward to explain the peculiar behavior of KIC 8462852, as outlined by \citet{Boyajian+16} and \citet{Wright&Sigurdsson16}.  These include stochastic variability associated with the activity of a very young star; obscuration by interstellar dust; occultation by debris from collisions between rocky bodies; extinction from periodic dusty outbursts of the R Coronae Borealis type; and transiting events due to a family of exo-comets or planetesimal fragments, either within the \tabby~planetary system itself (\citealt{Boyajian+16,Bodman&Quillen16}) or of an interstellar nature (\citealt{Makarov&Goldin16}).  However, other than the dips (and the peculiar secular dimming behavior discussed below), \tabby~appears typical in its properties, and there is no evidence$-$such as spectral emission lines or proper motion associated with a young cluster$-$to suggest a young age \citep{Boyajian+16}.  Furthermore, both R CrB and collision models should result in copious dust formation.\footnote{Indeed, some young stars, such as the AA Tau-like `dipper' systems (e.g.~\citealt{Ansdell+15}), which show sporadic minima of similar duration to \tabby, also show strong IR excesses and emission lines.}   However, \tabby~shows no persistent infrared (\citealt{Lisse+15,Marengo+15}) or sub-millimetre (\citealt{Thompson+16}) excess, ruling out the presence of significant quantities of long-lived dust, at least on small radial scales around the star (although one cannot yet exclude a mm or IR excess at times coincident with the transit events).  The observed photometric effects might be consistent with occultations from a giant artificial megastructure (\citealt{Wright+16}), although SETI optical (\citealt{Schuetz+16,Abeysekara+16}) and radio (\citealt{Harp+16}) searches have thus far been unsuccessful.

Adding to the mystery was the discovery by \citet{Schaefer16}, using the DASCH archival Harvard plates (\citealt{Grindlay+09}), that KIC 8462852 faded by 0.164 $\pm$ 0.013 magnitudes per century from 1890 to 1989, corresponding to a 14\% decrease in its luminosity.  \citet{Hippke+16a} argued that the accuracy of the digitized magnitudes of photometric plates on timescales of decades was insufficient to claim a significant detection of dimming (see also \citealt{Lund+16}; \citealt{Hippke+16b}b).  However, the star also dimmed by another 3\% over the 4.25 year duration of the {\it Kepler} mission \citep{Montet&Simon16}, with clear evidence that the dimming rate was not constant over this period.  Of a sample of 193 nearby comparison stars and 355 stars with similar stellar parameters, \citet{Montet&Simon16} find that none exhibit the same fading behavior as \tabby.

The transiting dips and secular dimming behavior are both unusual and hence by Ockham's razor are likely to be related phenomena.  However, they cannot both be readily explained as obscuration by the same cloud of material.  Analysis by \citet{Boyajian+16} shows that the transiting objects most likely originate within 10 AU of the star, whereas absorbing material on the same radial scale which continually covered $\sim 3-10\%$ of the stellar flux, as would be needed to explain the continuous dimming observed by \citet{Schaefer16} and \citet{Montet&Simon16}, would reradiate this emission above the mm/IR upper limits of $\sim 10^{-3}$ of the stellar luminosity.  Obscuration by interstellar clouds could evade these upper limits, provided that the clouds possessed both a smooth density profile to explain the secular dimming, as well as sub-AU structure to explain the dips \citep{Wright&Sigurdsson16}.

Here we consider an alternative explanation for the strange behavior of \tabby~as being the result of a possible tidal disruption and inspiral of a planetary body or series of planetary bodies into \tabby, which took place $\sim 10-10^{4}$ years ago, depending on the planet mass.  Gravitational energy released as a massive object sinks into the outer layers of the star causes a rapid and unobserved brightening, from which the stellar flux is now returning to the quiescent state, providing a possible explanation for the observed secular dimming of \tabby.  This possibility was first discussed by \citet{Wright&Sigurdsson16}.  Although they tentatively disfavored the idea due to the long photon diffusion time from the center of the star as being incompatible with the observed dimming timescale, they noted that a more detailed analysis of the hydrodynamics or stellar structure of a planet-star merger might reveal changes on faster timescales.  Indeed, as we show here, a range of dimming rates are achieved at different stages after such a merger event.

In our scenario, the occultation events could then be produced by the bound planetary debris from the same disruption event, or from a tidally-stripped moon system in the case of a massive planet.  Alternatively, the transits could arise from a group of comets or more massive bodies on high-eccentricity orbits, which are distinct from those responsible for the current secular dimming  (\citealt{Boyajian+16,Bodman&Quillen16,Budaj&Neslusan17}).  As we shall discuss, the hierarchical stellar binary architecture of the \tabby-system suggests that perturbations from the eccentric Lidov-Kozai mechanism could be driving a steady flux of planetary bodies into \tabby, more or less continually over its lifetime.  If this scenario is correct, our luck in catching \tabby~during its post-inspiral dimming therefore places stringent lower limits on the total planetary mass in the outer regions of F star systems.

This paper is organized as follows.  In $\S\ref{sec:Lidov-Kozai}$ we describe the Lidov-Kozai mechanism most likely responsible for driving a large flux of planetary bodies into \tabby.  In $\S\ref{sec:inspiral}$ we discuss our model for the luminosity evolution of \tabby~due to a planet inspiraling into the star, for different assumptions about the masses of the consumed planets.  In $\S\ref{sec:rates}$ we discuss constraints on our scenario based on the observed rate of \tabby-like phenomena within the {\it Kepler} field.  In $\S\ref{sec:transits}$ we discuss possible origins for the transiting clumps in our model.  In $\S\ref{sec:discussion}$ we provide a discussion and conclusion.  

\vspace{-0.5cm}

\section{The Eccentric Lidov-Kozai Mechanism}
\label{sec:Lidov-Kozai}

Consider a planet of mass $M_{\rm p} \ll M_{\star}$ orbiting \tabby~of mass $M_{\star}$, with a semi-major axis of $a_{\rm p} \gtrsim $ 10 AU, i.e. outside the ice line, corresponding to the extrasolar analog of the outer solar system or Kuiper belt.  Through either a strong scattering event with a massive planet or passing star (e.g.~\citealt{Rasio&Ford96}), or more gradually via secular processes, planetary bodies can be placed onto high eccentricity orbits with pericenter distances of $\sim 1-2R_{\odot}$, causing them to be tidally disrupted or directly consumed by the star.

\citet{Boyajian+16} detected an M-dwarf companion star (assumed mass $M_{\rm b} \simeq 0.4M_{\odot}$) at an angular distance of 1.96 arcsec from \tabby, corresponding to a physical distance of about $d_{\rm b} \approx 900$ AU if the M-dwarf resides at the same distance.  If the companion is bound in an orbit of semi-major axis $a_{\rm b} \sim d_{\rm b}$, then its orbital period would be $\sim 10^{4}$ yr.  One mechanism to drive planets to high eccentricity, over a timescale much longer than the orbital period, is the Lidov-Kozai mechanism (\citealt{Lidov62}; \citealt{Kozai62}).  The standard quadrupole-order Lidov-Kozai timescale (e.g.~\citealt{Liu+15}, their eq.~21)  is given by
\be
\tau_{\rm KL}^{\rm quad} \approx 5.3\,\,{\rm Myr}\left(\frac{a_{\rm p}}{20\,{\rm AU}}\right)^{-3/2}\left(\frac{M_{\rm b}}{0.4M_{\odot}}\right)^{-1}\left(\frac{M_{\star}}{1.43M_{\odot}}\right)^{1/2}\left(\frac{a_{\rm b}}{10^{3}{\rm AU}}\right)^{3}(1-e_{\rm b}^{2})^{3/2},
\ee
where $e_{\rm b}$ is the binary eccentricity and $M_{\star} \simeq 1.43M_{\odot}$ is the mass of \tabby. 

However, the fact that $\tau_{\rm KL}^{\rm quad}$ is much shorter than the age of \tabby~shows that the normal circular, quadrupole-order Lidov-Kozai mechanism cannot readily explain why planets in \tabby~would not have impacted the star much earlier in its evolution.  Furthermore, the inclination angle of the outer binary orbit would need to be fine-tuned to allow the pericenter radius of the planet to reach the very small values $\sim 10^{-2}$ AU required for direct interaction with \tabby.  

More promising is the eccentric Lidov-Kozai mechanism, which occurs when octupole contributions to the binary potential become important -  for example, when the outer binary has finite eccentricity ($e_{\rm b} \ne 0$).  This operates over a longer timescale and can reduce the planet pericenter to arbitrarily small values over timescales comparable to the stellar age (e.g., \citealt{Naoz+12,Li+14,Hamers+16}, see \citealt{Naoz16} for a recent review).\footnote{In principle, the gravity of a massive inner planet could act to ``detune" the Lidov-Kozai mechanism.  However, \citet{Boyajian+16} rule out the presence of a planet more massive than 20$M_{\rm J}$ with an orbital period shorter than a few hundred days.}  The octupole-order Lidov-Kozai cycle operates over a timescale which is approximately given by $\tau_{\rm KL}^{\rm oct} \sim \tau_{\rm KL}^{\rm quad}/\epsilon_{\rm oct}^{1/2}$ \citep[note that this differs from the value given in \citealt{Liu+15}]{Antognini15} where the dimensionless octupole moment is
\begin{equation}
\epsilon_{\rm oct} = \frac{M_\star - M_{\rm p}}{M_\star + M_{\rm p}} \frac{a_{\rm p}}{a_{\rm b}} \frac{e_{\rm b}}{1-e_{\rm b}^2}
\end{equation}
(e.g.~\citealt{Liu+15}), so that
\begin{eqnarray}
\tau_{\rm KL}^{\rm oct} &\simeq& \tau_{\rm KL}^{\rm quad}\sqrt{\frac{1-e_{\rm b}^{2}}{e_{\rm b}}\frac{a_{\rm b}}{a_{\rm p}}} \approx 4\times 10^7\,\,{\rm yr}\times \nonumber \\
&&\left(\frac{a_{\rm p}}{20\,{\rm AU}}\right)^{-2}\left(\frac{a_{\rm b}}{10^{3}{\rm AU}}\right)^{7/2}\left(\frac{M_{\rm b}}{0.4M_{\odot}}\right)^{-1}\left(\frac{M_{\star}}{1.43M_{\odot}}\right)^{1/2}\frac{(1-e_{\rm b}^{2})^{2}}{e_{\rm b}^{1/2}} \nonumber \\
\label{eq:tauKL}
\end{eqnarray}
If the observed companion M dwarf is on a moderately eccentric bound orbit with semimajor axis of $1800~{\rm AU}$ (higher than the observed separation due to projection effects) then $\tau_{\rm KL}^{\rm oct}$ can become comparable to the F star lifetime of $2$ Gyr.  This implies that over the lifetime of \tabby, the outer binary would ``drain" the inner stellar system of all planetary or planetesimal-sized bodies on radial scales of $a_p \gtrsim $ 10 AU, systematically driving them into \tabby.  

Excursions to very high eccentricity are possible, and in the portion of parameter space where orbit flips of point particles occur, maximum eccentricities of $e_{\rm max} \sim \sqrt{1-\epsilon_{\rm oct}^2}$ will be achieved \citep{Katz+11}.  For fiducial parameter choices ($a_{\rm p} \sim 10~{\rm AU}$, $a_{\rm b} = 1800~{\rm AU}$), then $a_{\rm p}(1-e_{\rm max}) << R_\odot$, and point particle orbits that nominally would flip instead collide with the star.  The fraction of parameter space where these high eccentricity excursions occur is difficult to quantify, but is fairly small for $\epsilon_{\rm oct} \lesssim 10^{-2}$ \citep{Antognini15}.

An additional constraint on the Lidov-Kozai mechanism is the requirement to avoid detuning by general relativistic precession.  This requires that the ratio of the semi-major axis of the inner binary to that of the outer one must obey (e.g.~\citealt{Blaes+02}, their eq.~A6; see also \citealt{Naoz+13})
\begin{eqnarray}
\frac{a_{\rm p}}{a_{\rm b}} &\gtrsim& \left(\frac{4 G(M_{\star}+M_{\rm p})^{2}(1-e_{\rm b}^{2})^{3/2}}{3 c^{2}M_{\rm b}a_{\rm p}(1-e_{\rm p}^{2})^{3/2}}\right)^{1/3} \nonumber \\
 &\approx&
  0.04\left(\frac{a_{\rm p}}{20\,\rm AU}\right)^{1/6}\left(\frac{r_{\rm p}}{4R_{\odot}}\right)^{-1/2}(1-e_{\rm b}^{2})^{1/2},
\end{eqnarray}
where in the second line we have taken $M_{\star} = 1.43M_{\odot}$, $M_{\rm b} = 0.4M_{\odot}$.  We have also used the fact that this requirement becomes most constraining near the end of the process, when the pericenter radius of the planet $r_{\rm p}$ is finally being driven close to the stellar surface, in which case $(1-e_{\rm p}^2)  \approx 2r_{\rm p}/a_{\rm p}$.  Therefore, for $a_{\rm b} \sim 10^{3}$ AU we require a semi-major axis for the planet of $a_{\rm p} \gtrsim 30-40$ AU.  Give how close this is to our fiducial value, a more detailed study is required to assess the quantitative impact of GR precession on the Lidov-Kozai process.

We explore several scenarios for the type of consumed planet or planets responsible for the irregular behavior of \tabby.  First, we consider the partial tidal disruption of a moon- or Earth-mass planet, in which the low density outer mantle of the planet (dominated by rock or ice) is violently stripped by stellar tides (e.g.~\citealt{Liu+13,Guillochon&RamirezRuiz13}), placing the debris on a range of eccentric orbits (see $\S\ref{sec:partial}$).  The remaining rocky or metallic core of the planet, itself denser than the star, subsequently impacts and spirals into \tabby.  In a second scenario, a Jupiter-mass planet or brown dwarf has its moon system tidally stripped before being consumed by the star.  Low mass bodies or stripped moons produced by the tidal disruption will have their original orbits strongly perturbed, and many will be shifted onto much more tightly bound orbits less vulnerable to the eccentric Lidov-Kozai mechanism.  These relatively stable orbits can provide long-term transiting events, as we describe in \S\ref{sec:transits}.  By contrast, the high mass planet or planetary core will continue on a similar orbit\footnote{The surviving core in the partial disruption can receive a kick velocity up to a magnitude comparable to the escape speed from the planet surface (\citealt{Manukian+13}), which in principle can exceed the orbital energy for initial semi-major axes $a_{\rm p} \gtrsim 10$ AU.  However, these large kicks only occur if an order unity fraction of the planetary mass is removed, and the core kick will be much less if the surviving core constitutes a large majority of the planet’s initial mass.}, continuing to experience eccentricity pumping and eventually coming to impact the star.  

Tidal disruption occurs once the pericenter radius of the planet $r_{\rm p}$ becomes less the tidal radius $R_{\rm t} \approx (1-2) R_{\odot}$, the precise value of which depends on the density of the planet (tidal stripping of moons will occur for somewhat larger pericenters).  How quickly $r_{\rm p}$ approaches $R_{\rm t}$ will depend on the mechanism responsible for pumping the planet's eccentricity.  In the quadrupole approximation, the maximum fractional change in pericenter radius $r_{\rm p} \approx j^{2}/(2GM_{\star})$ per orbit due to the eccentric Lidov-Kozai mechanism is given by (\citealt{Katz&Dong12}, their eq.~6),
\begin{eqnarray}
&&\frac{(\Delta j_{\rm max})^{2}}{j^{2}} = \frac{225\pi^{2}}{8}\left(\frac{M_{\rm b}}{M_{\star}}\right)^{2}\left(\frac{M_{\rm p}}{M_{\star}}\right)\left(\frac{a_{\rm p}^{7}}{a_b^{6} r_{\rm p}}\right) \nonumber \\
&&\approx 2\times 10^{-11}\left(\frac{M_{\rm p}}{M_{\rm J}}\right)\left(\frac{a_{\rm p}}{10{\rm AU}}\right)^{7}\left(\frac{r_{\rm p}}{2 R_{\odot}}\right)^{-1}\left(\frac{a_{\rm b}}{10^{3}\,{\rm AU}}\right)^{-6}\left(\frac{M_{\star}}{1.42M_{\odot}}\right)^{-3}\left(\frac{M_{\rm b}}{0.4M_{\odot}}\right)^{2}  \nonumber \\
\label{eq:jmax}
\end{eqnarray}
As long as $a_{\rm p} \lesssim 100$ AU we have $\Delta j_{\rm max}^{2} \ll j^{2}$, implying that many orbits are required to reduce $r_{\rm p}$ from $R_{\rm t} \gtrsim R_{\star}$ to $R_{\star}$.  This implies that the outer layers of the planet, or its moon system, would necessarily be removed in a sequence of partial disruption events before the remaining core impacts the stellar surface.  The stellar impact would also be a grazing one, such that the interaction is better described as a slow inspiral of the planet than a head-on collision ($\S\ref{sec:inspiral}$).

Tides can become important near pericenter for $r_{\rm p} \lesssim$ few $R_{\star}$, acting to circularize the planet's orbit before it reaches the tidal disruption radius (e.g.~\citealt{Fabrycky&Tremaine07}).  Although the rate of tidal circularization is uncertain for highly eccentric orbits, tides are also less likely to circularize low-mass rocky moons or Earth-mass planets than gaseous giants like Jupiter, due to the strong dependence of the tidal dissipation rate on the planetary radius.   

Finally, the eccentric Lidov-Kozai mechanism will result in the planet or cometary bodies which are being driven into \tabby~to arrive on orbits with a wide range of inclinations relative to the original orbital plane of the planetary system.  As we shall discuss, this could help explain the high occurrence rate of \tabby-like stellar systems \citep{Lacki16} because a favorable viewing angle is then not necessarily required to observe transiting events.  

\vspace{-0.4cm}

\section{Stellar Evolution Following Planet Inspiral}
\label{sec:inspiral}

In this section we quantify the change to the stellar luminosity following the inspiral of a planetary mass object into the core of an F-type main sequence star using the MESA\footnote{http://mesa.sourceforge.net, version 8118; default options used unless otherwise noted.} \citep{paxt11,paxt13,paxt15a} stellar evolution code.  We begin with a $1.43 \, M_\odot$ star and evolve it along the main sequence until it reaches a luminosity of $4.68 \, L_\odot$ \citep{Boyajian+16}, which occurs after $1.0$ Gyr.  A planet or planetary core of mass $M_p$ in a circular Keplerian orbit that has spiraled in to a radius $r$ within the star has a kinetic energy
\be
	E_{\rm kinetic}(r) = \frac{G M(r) M_p}{2 r} ,
\ee
where $M(r)$ is the spherically enclosed mass.  It feels an inwards gravitational acceleration of $G M(r)/r^2$, so that spiraling in a radial distance $dr$ implies  $GM(r)M_pdr/r^2$ worth of work has been done on it.  The work done on the object minus the change in the kinetic energy moving from $r$ to $r-dr$ is the energy that is deposited into the star.  We assume that this energy is locally converted into heat that is shared throughout the spherical shell, yielding a specific energy deposition
\be
	u_{\rm heat}(r) =  \frac{ GM(r)M_p}{8 \pi r^4 \rho(r) } + \frac{GM_p }{2r}  .
	\label{eq:uheat}
\ee

We further assume the object is tidally disrupted within the star at the radius, $r_{\rm disrupt}$, where the object fills its Hill sphere.  This condition is written $\rho_p = 9 M(r_{\rm disrupt}) / 4 \pi r_{\rm disrupt}^3$, where $\rho_p$ is the density of the object.  At this disruption radius, we assume the kinetic energy of the object is deposited as additional heat into a small region within a few numerical zones above $r_{\rm disrupt}$, which is equivalent to assuming the scale height at this depth is larger than the object.  This is only marginally true for Jupiter, but we neglect this effect for simplicity.  Since the inspiral is expected to occur rapidly, we neglect evaporation of the planet, which however could be more important for lower mass planetary bodies, or during the slower process of planetary inspiral into a red giant star (e.g.~\citealt{Livio&Soker84}).  

\begin{figure}
  \centering
  \includegraphics[width=1.0\columnwidth]{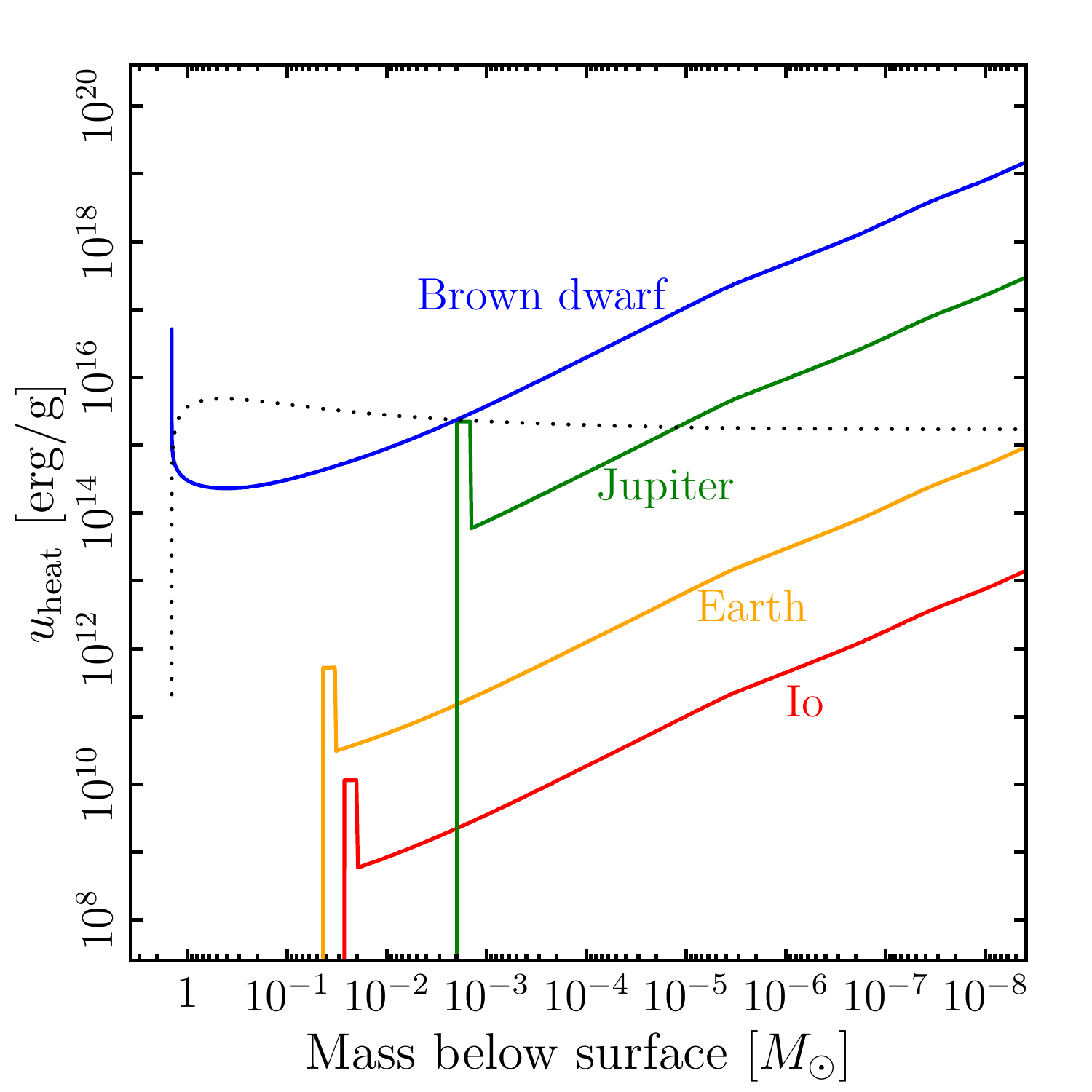}
  \caption{Heat deposited by the inspiral and tidal disruption of a planetary body, as a function of mass depth below the stellar surface, in our four fiducial scenarios (Table \ref{table:1}).  Shown with a dotted line is the specific gravitational binding energy $GM(r)/r$.}
  \label{fig:uheatvsm}
\end{figure}

As summarized in Table \ref{table:1}, we consider four fiducial cases: a moon of mass similar to Io or the Moon ($M_p=8.9 \times 10^{25}$ g, $\rho_p=10$ g cm$^{-3}$, assuming an Earth-like central density), Earth ($M_p=6.0 \times 10^{27}$ g, $\rho_p=13$ g cm$^{-3}$), Jupiter ($M_p=1.9 \times 10^{30}$ g, $\rho_p=4.0$ g cm$^{-3}$; \citealt{hubbard16}), and a brown dwarf 50 times more massive than Jupiter ($M_p=9.5 \times 10^{31}$ g, $\rho_p=660$ g cm$^{-3}$; \citealt{chabrier00}).  We use the central density of the object for $\rho_p$ in calculating the disruption radius $r_{\rm disrupt}$, except for Jupiter, for which we use an estimate for the density of the gaseous envelope just outside the rocky core because there is relatively little mass within the core.  The brown dwarf has a high enough density  that it reaches the center of the star without tidally disrupting.   Its specific heat deposition given by equation (\ref{eq:uheat}) approaches infinity at the center, but the integrated energy deposition is finite.  Additionally, its kinetic energy when it reaches the center approaches zero, so we do not add any additional heat.  The resulting specific heat deposition profiles are shown in Figure \ref{fig:uheatvsm}.  The dotted line is the specific gravitational binding energy $GM(r)/r$.

We implement these heating prescriptions into our MESA model by first reducing the maximum timestep to $10^3 \, {\rm s}$ and then depositing the specific energy from equation (\ref{eq:uheat}) over a duration of $10^4 \, {\rm s}$ for our four fiducial cases.  As the dotted line in Figure \ref{fig:uheatvsm} shows, the energy deposition in the outer layers is enough to overcome the gravitational binding energy for some of the models, so we implement a Roche lobe overflow prescription that removes mass once it expands past $2 \, R_\odot$.

\begin{figure}
  \centering
  \includegraphics[width=1.0\columnwidth]{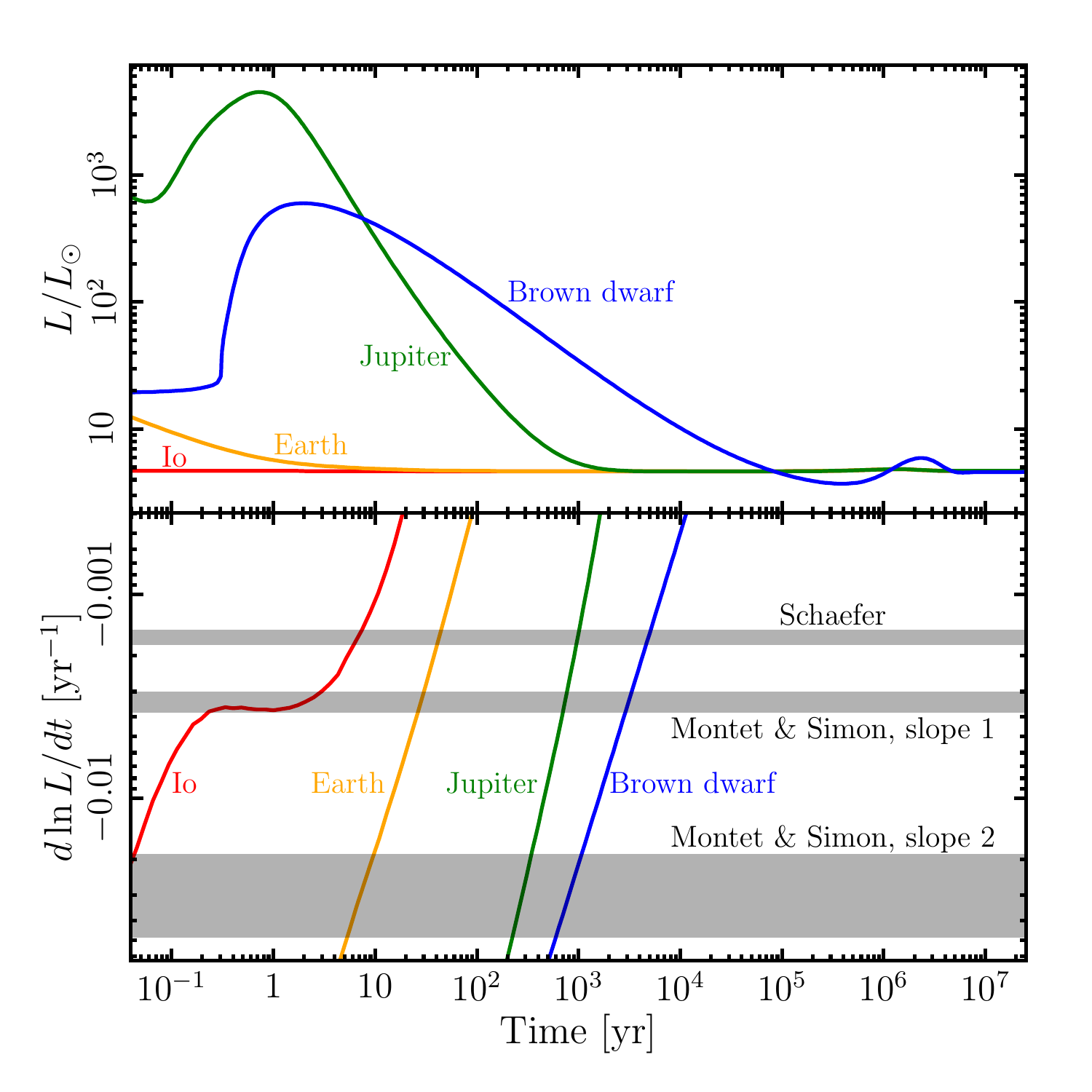}
  \caption{The top panel shows the stellar luminosity as a function of time following a planetary inspiral, as calculated by MESA, in our four fiducial scenarios (Table \ref{table:1}).  The bottom panel shows the rate of stellar dimming d$lnL$/d$t$ for each model, in comparison to the inferred rate of dimming from \citet{Schaefer16} and \citet{Montet&Simon16}.  In the latter case, we show the dimming rate both before and after the acceleration in the dimming rate.  }
  \label{fig:lvst}
\end{figure}

\begin{figure}
  \centering
  \includegraphics[width=1.0\columnwidth]{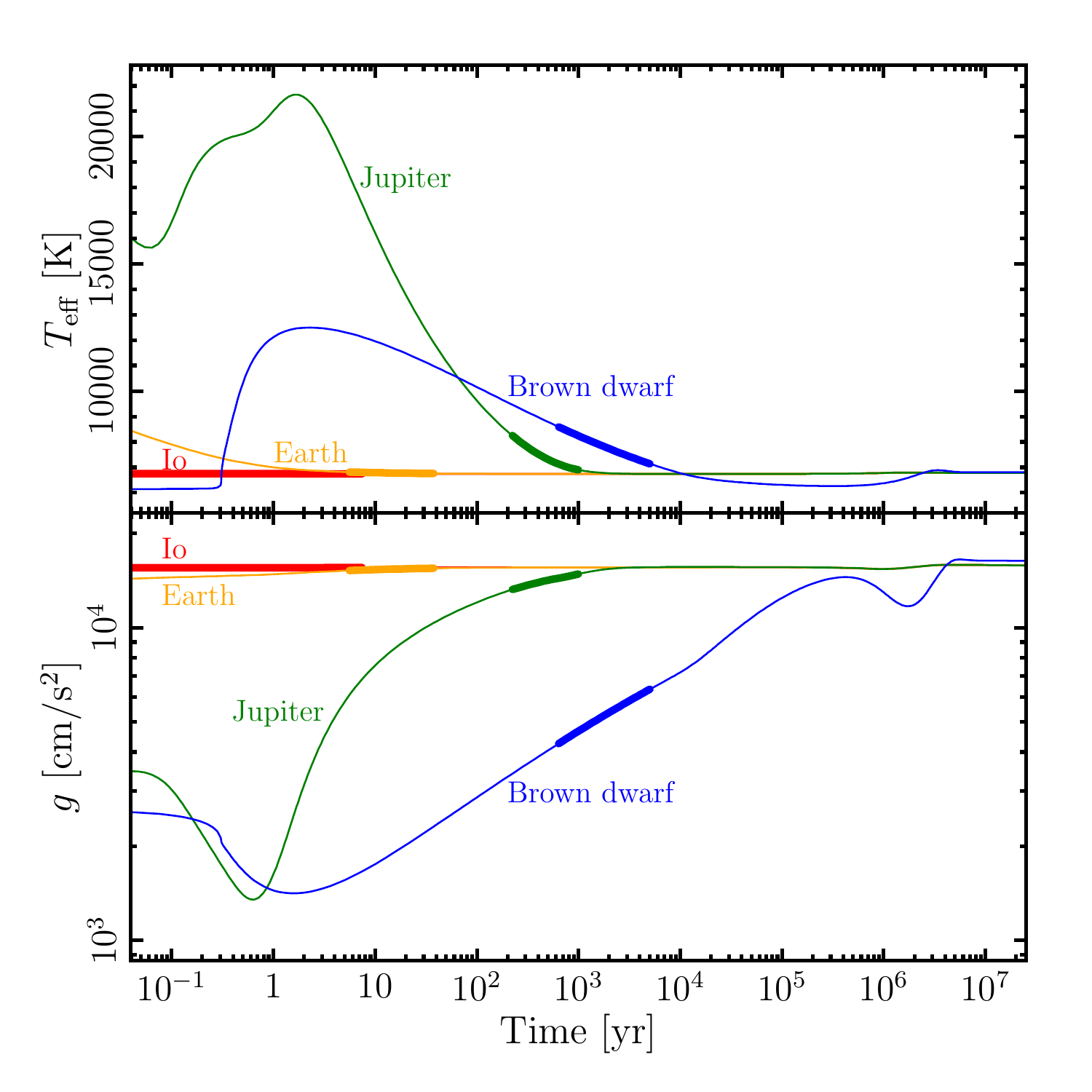}
  \caption{Surface gravity and effective temperature for the same models shown in Figure \ref{fig:lvst}.  Thickened lines denote the epoch over which the dimming rate lies between the \citet{Schaefer16} and \citet{Montet&Simon16} values (including error bars).}
  \label{fig:gteffvst}
\end{figure}

The resulting light curves are shown in the top panel of Figure \ref{fig:lvst}, and the normalized rate of dimming, d $\ln L$/d$t$, is shown in the bottom panel.  All four models exhibit a quick rise to a peak luminosity, followed by a longer phase of dimming.  Also shown as bands in the bottom panel are the average rate of dimming reported by \citet{Schaefer16} from photographic plates over a span of $100 \, {\rm yr}$ and by \citet{Montet&Simon16} from an examination of {\it Kepler} data within a recent $4$-yr period.  In the latter case, we show two dimming rates, corresponding to that observed before and after, respectively, the acceleration in the dimming rate observed midway through the mission.  Due to the varying amount and depth of the energy deposition, the four models dim at different rates for different periods of time: Io matches the observational rate of dimming for $\simeq 5 \, {\rm yr}$, Earth for $\simeq 30 \, {\rm yr}$, Jupiter for $\simeq 800 \, {\rm yr}$, and a 50 Jupiter-mass brown dwarf for $\simeq 4000 \, {\rm yr}$.  These numbers are summarized in Table \ref{table:1}.  

In order to explain the \citet{Montet&Simon16} dimming rate alone, one can therefore invoke the consumption by \tabby~of a moon-mass body roughly contemporaneous with the beginning of the {\it Kepler} mission.  However, this would probably be insufficient to explain the longer timescale dimming observed by \citet{Schaefer16}, which would instead require a body comparable or more massive than the Earth being consumed by \tabby~on a timescale of $\gtrsim 10^{2}$ yr ago.  One speculative unification possibility is that a planet with mass comparable or greater than Earth was disrupted $\gtrsim 200$ years ago, giving rise to the \citet{Schaefer16} dimming, followed after several additional orbits by the consumption of its tidally-stripped moon roughly a decade ago, explaining the \citet{Montet&Simon16} dimming.  

Figure \ref{fig:gteffvst} shows the effective temperature (top panel) and surface gravity (bottom panel) of the same models shown in Figure \ref{fig:lvst}.  We observe that the increase in stellar luminosity following planet consumption generally results from a combination of higher effective temperature and larger stellar radius (lower gravity).  Thickened lines denote the epoch over which the dimming rate lies between the \citet{Schaefer16} and \citet{Montet&Simon16} values (including error bars).  Although the effective gravity of the brown dwarf case appears inconsistent with the measured value of log$(g) = 4.0 \pm 0.2$ for \tabby~(\citealt{Boyajian+16}), this is somewhat deceptive because our stellar evolution models have assumed a value of $1.43M_{\odot}$ for the mass of \tabby: if we had instead assumed a less massive star with an initially high surface gravity, it is possible that a brown dwarf merger model could be found which matches the observed rate of dimming as well as the current gravity and effective temperature of \tabby.  

\vspace{-0.5cm}

\section{Rate Constraints}
\label{sec:rates}

Although \tabby~is unique in terms of its transiting behavior within the {\it Kepler} database (\citealt{Boyajian+16}), on the whole such behavior must be surprisingly common (\citealt{Lacki16}).  There are about $N_{\rm F} = 5000$ F stars similar to \tabby~($T_{\rm eff} = 6750$ K) in the {\it Kepler} field.  An F star main sequence lifetime of $t_{\rm F} = 2\times 10^{9}$ yr would thus require an ``on"-time for \tabby-like phenomena of $t_{\rm on} = (t_{\rm F}/N_{\rm F})f_{\rm t}^{-1} \simeq 4\times 10^{5}f_{\rm t}^{-1}$ yr, where $f_{\rm t} < 1$ is the fraction of the total solid angle within which an external observer would observe the transits.  A quantitatively similar on-time is obtained by broadening consideration to the larger number $\sim 10^{4.5}$ of solar-type stars in the {\it Kepler} field, given also their longer lifetimes of (so far) $\sim 10^{10}$ yr.  These sizable on-times imply that a large quantity of mass in planets must be consumed by a typical F star over its lifetime to explain the secular dimming of \tabby.  

We first consider a semi-model-independent constraint on the minimum required planetary mass.  After the star consumes a total mass $M_{\rm tot}$ of planetary bodies, the maximum enhancement to its total radiated energy is roughly given by $E_{\rm rad} \sim GM_{\rm tot}M_{\star}/2R_{\star}$.  In order to explain the fractional flux change observed by \citet{Schaefer16} or \citet{Montet&Simon16} of $f_{\rm L} \gtrsim 0.03-0.1$, the required rate of energy release over the lifetime of a typical F star is at least $f_{\rm L}L_{\star}t_{\rm on}$, under the assumption that all F stars go through a similar phase of planetesimal consumption, where $L_{\star} \simeq 4.7L_{\odot}$ is the unperturbed luminosity of \tabby.  Equating this to $E_{\rm rad}$ we obtain the minimum mass of consumed planets by a typical F star in order to explain the \tabby~dimming phenomenon,
\be
M_{\rm tot} \gtrsim M_{\rm min} \simeq \frac{2f_{\rm L}L_{\star}t_{\rm on}R_{\star}}{GM_{\star}} \approx 3\times 10^{3}M_{\oplus}f_{\rm t}^{-1}\left(\frac{f_{\rm L}}{0.1}\right).
\label{eq:Mmin}
\ee    
Even if transiting debris covers most observer viewing angles (e.g.~$f_{\rm t} \sim 1$ due to the range of inclinations induced by eccentric Lidov-Kozai oscillations; $\S\ref{sec:Lidov-Kozai}$), a typical F star must consume 10 Jupiter masses over its lifetime.  

In fact, the minimum mass $M_{\rm min}$ given by equation (\ref{eq:Mmin}) strictly applies only if the planetary bodies are moons or low mass planets, for which the observed dimming time is $t_{\rm dim} \sim 10$ years and the peak luminosity following a single consumed object $E_{\rm rad}/t_{\rm dim} \sim 0.1L_{\star}$.  For more massive planets like Jupiter the peak luminosity achieved by the star substantially exceeds a 10\% excess and hence the required minimum mass $M_{\rm min}$ is larger by 1-2 orders of magnitude.  In Table \ref{table:1} we compile the true minimum total number (and corresponding minimum mass) for each type of consumed body, which we have estimated as
\be
N_{\rm min} = N_{\rm F}\left(\frac{t_{\rm F}}{t_{\rm dim}}\right),
\ee
where $t_{\rm dim}$ is the timescale from our MESA calculations over which the dimming rate of the stellar luminosity for each body is $dln L/dt \gtrsim 0.0025$ yr$^{-1}$, a number midway between the \cite{Schaefer16} and \citet{Montet&Simon16} dimming rates.

On the face of it, the required planetary masses of $\sim 10^{3}-10^{6}M_{\oplus} \sim 3-3000M_{\rm J}$ per star appears implausibly high.  This is especially true considering that the outer binary architecture of the \tabby~system, which may give rise to a flux of planets into the central star through the eccentric Lidov-Kozai mechanism ($\S\ref{sec:Lidov-Kozai}$), is not generic and hence the required mass in the \tabby~system would be even higher.  On the other hand, the statistics of one system do allow for \tabby~to be an outlier, in particular when accounting for the `look-elsewhere' effect that afflicts inferences in systems with a large parameter space of possible strange behaviors.  Our interpretation nevertheless would require that a large fraction of F stars contain favorable stellar binary companions and massive planetary systems, in order to prevent \tabby~from being a major statistical outlier.

Table \ref{table:1} makes clear that the most ``economical'' approach in terms of minimizing the total required planetary mass is to invoke a large number of moon-mass objects, the same size scale required to explain the fast variability timescale observed by \citet{Montet&Simon16}.  Such objects are comparable in size to the pre-planetary ``embryos" created during the planet formation process \citep{Armitage10}.  A high inefficiency in transforming such embryos into planets could have left the required large reservoir of such objects in the outer regions of a planetary system like that of \tabby.\footnote{We thank the anonymous reviewer for pointing out this possible connection.}  However, we note that even in this scenario the minimum required mass of $3 M_{\rm J}$ is a sizable fraction of the total mass of metals (non H or He) in \tabby.  

While the rate at which stars more massive than the Sun host outer planetary systems is uncertain due to observational challenges (e.g.~\citealt{Nielsen+13}), A star systems can possess massive debris disks (e.g.~\citealt{Su+06}).  For instance, \citet{Thureau+14} find between $10^{-7}-0.1M_{\oplus}$ in mm to cm size dust.  If we extend this distribution to higher mass objects using $dN/da \propto a^{-3.5}$ (where $a$ is the particle radius and 3.5 is the \citet{Dohnanyi69} exponent for a collisional cascade), this corresponds to a total debris disk mass of $10^{-3}-3000M_{\oplus}$ if we truncate the upper bound of the cascade at 1000 km scale objects similar to the Moon or Io.  This very upper range is consistent, albeit barely, with the required mass of Io-size bodies per star from Table \ref{table:1}.

\section{Origin of the Transiting Clumps}
\label{sec:transits}

The transiting `clumps' observed by {\it {\it Kepler}} produce flux deficit percentages from $f = 0.005-0.2$ and  typically last a few days \citep{Boyajian+16}.  The flux deficit percentages require obscuring `clouds' with surface areas $\Sigma \gtrsim f\pi R_{\star}^{2}$ and characteristic radii of $R_{\rm c} \gtrsim f^{1/2}R_{\star} \sim 0.1-1R_{\odot} $; the cloud radius exceeds this minimum value if it is not optically-thick.  Such large clouds are obviously not gravitationally bound objects, but they could be clouds of gas or dust surrounding planetesimal-mass bodies \citep{Boyajian+16,Budaj&Neslusan17}.  In this section we outline several possible explanations for the underlying bodies responsible for these transiting events within our proposed scenario, all of which are causally connected to our model for the secular dimming following planet impact with the star.        

\subsection{Scenario 1: Comet- or planetesimal-mass bodies not formed from planet-star collision}
\label{sec:comet}

One possibility is that the transits are due to out-gassing from a swarm of comet- or planetesimal-mass bodies on high eccentricity orbits \citep{Boyajian+16,Bodman&Quillen16}, which are not directly a result of the planet-star collision responsible for the secular dimming.  In our scenario for \tabby, a steady flux of low-mass bodies into the inner stellar system would be naturally predicted as resulting from the same mechanism  responsible for driving larger mass moons or planets into the star ($\S\ref{sec:Lidov-Kozai}$).  As was already discussed, such bodies could occupy a wide range of orbital inclinations due to the Lidov-Kozai process, helping to relax constraints on the the high occurrence rate of \tabby-like systems.  \citet{Bodman&Quillen16} suggest a single comet family from a tidally disrupted Ceres-sized progenitor or the start of a Late Heavy Bombardment period can explain the second major structure (D1500) in the \tabby~light curve.  However, they could not produce the deepest D800 event with a comet-like structure.  

The asymmetric shape of the transits suggest that the obscuring material could well be highly elongated.  \citet{Boyajian+16} note a possible tension with the usual cometary head/lagging tail hypothesis, based on the ingress being slower than egress in some of the dips (see also \citealt{Budaj&Neslusan17}).  Such a geometry might be more easily explained if the obscuring material is undergoing a Roche-lobe overflow near pericenter, in which case the bodies responsible for the obscuration could be much more massive than typical comets.  \citet{Budaj&Neslusan17} explored the orbital evolution behavior of initially spherical dust cloud around central objects of mass ranging from large cometary nuclei ($\approx 10^{-6}M_{\oplus}$) to a large moon ($\approx 5\times 10^{-2}M_{\oplus}$), which they showed can indeed produce transiting dips in agreement with those of \tabby~using only four bodies shrouded in dust clouds.  

 A potential challenge with the ``unrelated swarm" hypothesis is that the inspiral of a massive body into the central star is expected to be rare compared to lower-mass bodies capable of producing the observed dimming (which, of course, are what led to \tabby~being flagged as interesting in the first place). 
In the next two sections we discuss ideas for producing the obscuring bodies directly from the same planetary disruption process responsible for the secular dimming.

\begin{deluxetable}{lcccc}
\tablenum{1}
\tabletypesize{\normalsize}
\tablecolumns{5}
\tabcolsep0.05in\footnotesize
\tablewidth{0pt}
\tablecaption{ \label{table:1}}
\tablehead {
\colhead{Body}                   &
\colhead{$M_p$}  &
\colhead{$t_{\rm dim}^{(a)}$}           &
\colhead{$M_{\rm min}^{(b)}$}    &
 \colhead{$N_{\rm min}^{(c)}$} \\
\colhead{}  &
\colhead{($M_{\oplus}$)}      &
\colhead{(yr)}       &
 \colhead{($M_{\oplus}$)} & 
 \colhead{} 
}
\startdata
Moon (Io) & 0.015 & 4.5 & $1.3\times 10^{3}$ & $9\times 10^{4}$ \\
Earth & 1 & 35 & $1.1\times 10^{4}$ & $1.1\times 10^{4}$ \\
Jupiter & 318 & 850 & $1.5\times 10^{5}$ & 470 \\
Brown  Dwarf & $1.6\times 10^{4}$ & $4\times 10^{3}$ & $1.6\times 10^{6}$ & 100 \\
\enddata
\begin{minipage}{\columnwidth}
\tablecomments{$^{(a)}$Dimming timescale over which the dimming rate of the stellar luminosity is $dln L/dt \gtrsim 0.0025$ yr$^{-1}$, a number midway between the \cite{Schaefer16} and \citet{Montet&Simon16} dimming rates.  $^{(b)}$Total mass consumed in bodies of mass $M_p$ to explain inferred rate of \tabby-like stars assuming transits cover all solid angles ($f_t = 1$). $^{(c)}$Total number of bodies consumed to explain inferred rate of \tabby-like stars assuming transits cover all solid angles ($f_t = 1$).}
\end{minipage}
\end{deluxetable}

\subsection{Scenario 2: Partial Disruption of a Differentiated Planet}
\label{sec:partial}

An Earth-mass planet could experience a partial disruption of its outer layers prior to the consumption of its core by a direct impact and inspiral into \tabby.  The tidal disruption of a planet induces a dispersion in the orbital properties of the resulting debris, which can be computed assuming a relatively impulsive disruption, i.e. that at the moment the planet crosses into the tidal sphere, it shatters into constituent pieces that retain the center of mass velocity but now occupy a range of spatial coordinates.  The consequent spread in specific binding energy of the disrupted debris is approximately
\be
\Delta \epsilon = \frac{GM_{\star}R_{\rm p}}{R_{\rm t}^{2}},
\label{eq:deltae}
\ee
where again $R_{\rm t}$ is the tidal radius.  The ratio of this energy to the original specific gravitational binding energy of the planet orbit, $\epsilon_{\rm p} = GM_{\star}/2a_{\rm p}$ is given by
\be
\frac{\Delta \epsilon }{\epsilon_{\rm p}} \approx \frac{2 R_{\rm p}a_{\rm p}}{R_{\rm t}^{2}} \approx 10\left(\frac{a_{\rm p}}{10{\rm AU}}\right)\left(\frac{R_{\rm p}}{R_{\oplus}}\right)\left(\frac{R_{\rm t}}{2R_{\odot}}\right)^{-2}
\ee
The fact that $\Delta \epsilon/\epsilon_{\rm p} \gg 1$ indicates that roughly half of the debris will be unbound from the system, while the other half will be placed onto tighter orbits with orbital energy $-\Delta \epsilon$, i.e. with characteristic semi-major axes of
\be
a_{\rm deb} \simeq \frac{R_{\rm t}^{2}}{R_{\rm p}} \approx 2{\rm AU}\left(\frac{R_{\rm t}}{2R_{\odot}}\right)^{2}\left(\frac{R_{\rm p}}{R_{\oplus}}\right)^{-1}
\label{eq:adeb}
\ee  
and orbital periods of
\be
T_{\rm deb} \approx 2.4\,\,{\rm  yr}\left(\frac{R_{\rm t}}{2R_{\odot}}\right)^{3}\left(\frac{R_{\rm p}}{R_{\oplus}}\right)^{-3/2}
\label{eq:Tdeb}
\ee
Because $a_{\rm deb}$ is generally less than $a_{\rm p}$, orbital perturbations from the outer binary will have much less effect on the debris streams than they will on the surviving core of the partially disrupted planet, which to first order retains its original semimajor axis.  The transiting debris should therefore remain ``frozen in'' to its orbits for much longer than it takes the surviving planetary core to be ingested by the star.

The partially disrupted debris streams may recollapse in the transverse direction under the influence of stream self-gravity, possibly even fragmenting (via a sausage instability) into completely self-bound clumps of the type necessary to produce discrete transiting events.  This gravitationally-induced clumping has been observed to occur in simulations of stellar tidal disruption (e.g.~\citealt{Coughlin&Nixon15}), but as these simulations focus on the full disruption regime, it is unclear how applicable they are to partial disruptions.  

Although the dipping behavior in \tabby~showed no clear periodic signals\footnote{
\citet{Boyajian+16} detected a 0.88 day periodicity, which they attributed as being likely due to the rotation period of KIC 8462852, as supported also by spectroscopy (with some lower frequency noise suggesting 10\% differential rotation).  However, \citet{Makarov&Goldin16} attribute this variability to a different star on the same {\it {\it Kepler}} channel. }, the main dip structures D800 and D1500 were separated by a timescale of 700 days, which is indeed comparable to $T_{\rm deb}$ (eq.~\ref{eq:Tdeb}) if the radius of the disrupted planet is comparable to that of the Earth.  In this scenario, a smaller fraction of the bound debris with energy between $-\Delta \epsilon$ and 0 will form a series of nested elliptical orbits with semi-major axes $a \gtrsim a_{\rm deb}$. 
Obscuration by this matter on orbits with $T \gtrsim T_{\rm deb}$ could explain dipping events at phases different from those of the main dips.


\subsection{Scenario 3: Tidally-Stripped Moon System}
\label{sec:moon}

If the recently ingested planetary object is a gas giant similar to those in the Solar System, it will have a system of exomoons surrounding it.  Because $a_{\rm p}(1-e_{\rm max}) \ll R_\odot$, the planet's pericenter will slowly diffuse inward toward the star, with ample opportunity for the moon system to be completely stripped before the planet impacts or inspirals into the star.  This will leave a chain of tidally detached exomoons on highly eccentric orbits around the host star, but whose semimajor axes are much smaller than the original semimajor axis of the disrupted planet: $a_{\rm moon}\approx \tilde{a}_{\rm moon}(M_\star / M_{\rm p})^{1/3}$ (where $\tilde{a}_{\rm moon}$ is the original semimajor axis of the moon's orbit about its planet).  Since $a_{\rm moon} \ll a_{\rm p}$, generally, the stripped exomoons' Kozai timescale $\tau_{\rm KL}^{\rm oct} \gg ~{\rm Gyr}$ and they will be dynamically stable to perturbations from the binary M star (as we argued in the prior subsection would be the case for partially stripped planetary debris).

After the host planet is ingested, the surviving exomoons will maintain their highly eccentric, low-pericenter orbits.  Near pericenter, they may receive significant heating both from tides and from irradiation.  If the cooling time of the exomoons is long compared to the orbital period, these periodic heating events will build up.  If the entire planet melts, it can begin catastrophic thermal atmosphere loss; gases will escape with a characteristic velocity $v_{\rm moon} \sim \sqrt{Gm_{\rm moon}/r_{\rm moon}}\sim 1~{\rm km~s}^{-1}$.  Gas lost thermally at apocenter will follow the moon's center of mass trajectory and likely remain bound to it in an extended cloud; gas lost thermally at pericenter will escape with an energy spread $\delta \epsilon \sim v_{\rm moon}\sqrt{GM_\star/r_{\rm p, moon}}$ (where $r_{\rm p,moon}$ is the pericenter radius of moon), which is also low enough to remain on the moon's orbit.

Alternatively, if the exomoon does not completely melt, tidal deposition of energy deep in the interior will cause volcanic out-gassing analogous to Jupiter's tidally-heated moon Io (e.g.~\citealt{McEwen&Soderblom83,Graps+00}).  The most common type of volcanic out-gassing on Io are dust plumes produced when encroaching lava flows vaporize underlying sulfur dioxide frost, sending the material skyward.  These plumes are usually less than 100 kilometres tall with eruption velocities around 0.5 km s$^{-1}$; however, the tidal heating in our scenario is much more extreme than in the Io-Jupiter system.  Both of these mechanisms may allow for the creation of large, extended clouds around a population of tidally detached exomoons.  These are potentially compatible with the four bodies \citet{Budaj&Neslusan17} find could explain the transits of \tabby.

\subsection{Constraints from Infrared/Millimetre Dust Emission}
\label{sec:dust}
 Another motivation for favoring eccentric bodies as the source of the obscuring material responsible for the transiting dips is that it becomes possible to evade tight upper limits on the persistent IR and mm emission, which are constrained to be $\lesssim 10^{-3}$ of the stellar luminosity, depending on the temperature of the material (\citealt{Lisse+15,Marengo+15,Thompson+16}).  A dust cloud with a fixed grain surface area capable of producing deep transits close to its pericenter passage would, if moving along the same orbit as the out-gassing body, cover a much smaller fraction of the solid angle of the star near apocenter (where it also spends most of its time).  Furthermore, given that in our partial tidal disruption ($\S\ref{sec:partial}$) and tidally-stripped exomoon ($\S\ref{sec:moon}$) scenarios the transiting debris has a pericenter radius of a few stellar radii, dust would be unlikely to survive more than one orbit due to sublimation near pericenter.

Still, because of the energy spread induced by the out-gassing processes, we expect that the dust would quickly become spread throughout the orbital phase.  This is especially true in the eccentric swarm picture described in $\S\ref{sec:comet}$, where, due to the larger pericenter radius, dust could survive several orbits without sublimating.  The most efficient mechanism for removing large grains is Poynting Robertson (PR) drag, which for a highly eccentric orbit of pericenter radius $r_{\rm p}$ and semi-major axis $a$ occurs on a timescale (\citealt{Stone+15}, their Appendix D)
\be
t_{\rm PR} \approx \frac{16 \sqrt{2}}{15}\frac{b \rho_{\rm d}c^{2} a^{1/2}r_{\rm p}^{3/2}}{L_{\star}} \approx 49\,{\rm day}\,\,\left(\frac{b}{1\,\mu{\rm m}}\right)\left(\frac{r_{\rm p}}{2R_{\odot}}\right)^{3/2}\left(\frac{a}{2\rm AU}\right)^{1/2},
\ee
where $b$ is the radius of the spherical grain, which we have assumed exceeds the wavelength of the stellar light, $L_{\star} = 4.7L_{\odot}$ is the stellar luminosity, and $\rho_{\rm d} = 2$ g cm$^{-3}$ is the assumed bulk grain density.  Small dust grains are removed much faster (on the local dynamical timescale or less) by radiation blow-out, as occurs for particles of size
\be
b \lesssim \left(\frac{3}{4\pi}\frac{L_{\star}}{GM_{\star}\rho_{\rm d}c}\right) \simeq 4\mu m,
\ee
where we have assumed $M_{\star} = 1.43M_{\odot}$.    

The characteristic size of dust grains is obviously uncertain in our various proposed scenarios, but it could be less than a few microns, i.e. within the blow-out regime.  The solid-state debris left over from cometary sublimation within our solar system possesses a size distribution $dn/db \propto (b/b_{\rm min})^{N}$ above a characteristic minimum size $b_{\rm min} \approx 0.1\mu$m, where $3.7\lesssim N \lesssim 4.3$ (e.g., \citealt{Harker+02}).\footnote{The {\it maximum} size of cometary debris is difficult to measure, although an analysis of measurements by NASA's Deep Impact mission to the comet 9P/Tempel 1 led to estimates of $1 \mu$m $\lesssim b_{\rm max} \lesssim 100\mu$m (\citealt{AHearn+05,Gicquel+12}).}  In our volcanic out-gassing scenario ($\S\ref{sec:moon}$), some guidance is provided from measurements by the Galileo satellite of the `Loki' plume on Io, which found it to be comprised mainly of $\sim 0.001-0.01 \mu$m (`smoke') particles (\citealt{Collins81,Ip96}).

Although the lack of a persistent mm/IR excess from reprocessed stellar light is not yet restrictive on our model, such emission should still accompany the transiting dips close to pericenter.  One way to test this hypothesis would be to obtain IR or mm observations {\it during} a transit event.  In our partial TDE ($\S\ref{sec:partial}$) and tidally-stripped exomoon ($\S\ref{sec:moon}$) scenarios, the bodies responsible for the transiting debris reach pericenter well within the dust sublimation radius of about $R_{\rm sub} = (T_{\rm eff}/T_{\rm sub})^{2}R_{\star} \approx 30 R_{\odot}$, where $T_{\rm eff} = 6750$ K and $R_{\star} = 1.58R_{\odot}$ are temperature and radius of \tabby and $T_{\rm sub} \simeq 1500$ K is the sublimation temperature of silicate grains.  If out-gassing is happening continuously, we should therefore expect the IR emission to grow in peak frequency and luminosity as the dusty cloud approaches $R_{\rm sub}$.  The IR emission will then subside for a time $t_{\rm off} = 2R_{\rm sub}/v_{\rm sub} \approx 3$ days during pericenter passage, where $v_{\rm sub} = (2GM_{\star}/R_{\rm sub})^{1/2}$ is velocity of the nearly parabolic orbit, before possibly resuming once the orbit again exits the sublimation zone (assuming out-gassing is still ongoing).  Depending on the phase of the orbit at which the transit dips occur, the centroid of this double-horned-shaped IR light curve could be offset in time by the dimming events by a few days or longer (before or after, depending on whether the transits occur during the ingoing or outgoing phase of the cloud orbit).

\section{Discussion and Conclusions}
\label{sec:discussion}

Building on an initial suggestion by \citet{Wright&Sigurdsson16}, we have considered an explanation for the secular dimming behavior in \tabby~observed by \citet{Schaefer16} and \citet{Montet&Simon16} as being due to a past inspiral of a moon- or planet mass body into the star.  As a proof of principle, we have calculated the time evolution of the luminosity of a main sequence F star (Fig.~\ref{fig:lvst}), including a simplified model for the gravitational energy released throughout the star from planetary inspiral (Fig.~\ref{fig:uheatvsm}), with MESA.  We find that the observed dimming behavior of \tabby~can be explained as the slow decline in luminosity on timescales of $\sim 10-10^{4}$ yr, depending on the mass of the consumed planet.  

Our simple model cannot readily explain the abrupt change in the photometric decay rate from slower to fast decline observed by \citet{Montet&Simon16}.  In principle such substructure could result from a more complicated radial energy-deposition profile in the star than we have assumed from the inspiraling planet, and/or if multiple fragments fell into the star over the {\it Kepler} mission.  However, given the many other uncertainties in our treatment (e.g. effects of rotation, our simplifed stellar mass loss prescription, our assumption that heat deposited by the planet is rapidly distributed over spherical shells), we relegate a more thorough study of this issue to future work.

Perhaps the biggest current challenge to our model is the high required occurrence rate of planet-star impact events, given the detection of even a single such dimming event in the {\it Kepler} sample (see also \citealt{Lacki16}).  Reducing the statistical improbability of observing a \tabby-like system to a reasonable value requires both a high total mass of planets in F star systems on radial scales $r \sim 1-100$ AU, and an efficient mechanism for driving them into the central star over a timescale comparable to the stellar lifetime.  For the latter, we have invoked the eccentric Lidov-Kozai mechanism ($\S\ref{sec:Lidov-Kozai}$), which  could be efficient at draining the planetary system into \tabby~if the apparent M-dwarf companion of \tabby~is indeed gravitationally bound to it.  

High mass debris disks around stars more massive than the Sun are indeed inferred around A stars  (e.g.~\citealt{Su+06}), slightly more massive than \tabby.  They are also inferred indirectly based on high rates of metal pollution onto their white dwarf by asteroid pollution events (e.g.~\citealt{Farihi16} and references therein).  N-body calculations based on an extrapolation of our own solar system architecture through the AGB and white dwarf phase suggest that explaining the inferred metal accretion rates of young white dwarfs require asteroid belts with masses exceeding those in our solar system by a factor of 1000 (e.g., \citealt{Debes+12}).   

Although in principle the dimming of \tabby~could be explained as the disruption of a Jupiter-mass planet $\sim 10^{4}$ years ago, for reasons of economy regarding the total required mass we are pushed towards instead invoking a large number of smaller objects, such as $10^{3}$ km-size moon-like bodies (Table~\ref{table:1}).  Such smaller objects might also produce a dimming rate consistent with the faster variability observed by \citet{Montet&Simon16}.  Considering the disruption of Earth-mass planets also makes the required masses uncomfortably large.  However, one benefit of invoking an Earth-mass disruption in \tabby~is that the bound debris from the partial disruption is predicted to occupy orbits with characteristic periods of a few years (eq.~\ref{eq:Tdeb}), comparable to the observed 800 day interval between the D800 and D1500 dip clusters (should this apparent periodicity be more firmly established by further monitoring of \tabby).

A strength of our model is that it provides a plausible causal connection between secular dimming and the short-timescale dipping behavior, which to our knowledge is absent from other proposed explanations.  A large flux of low-mass bodies into the inner stellar system - on a variety of orbital inclinations with respect to the original planetary disk - arises naturally from the same mechanism causing planet-star impacts.  One of the motivations for considering a swarm of {\it eccentric} bodies as the source of the obscuring material responsible for the dips is that it allows one to evade tight upper limits on the IR and mm flux due to reprocessed stellar luminosity (\citealt{Lisse+15,Marengo+15,Thompson+16}), which should nevertheless still accompany the transiting dips (which likely occur near the pericenter radii of the obcuring matter).  This hypothesis could be tested with IR or mm observations {\it during} an obscuration event ($\S\ref{sec:dust}$).  Indeed, flux dips up to $\sim 20\%$ should be readily detectable from ground-based monitoring, enabling such a triggered observational programme.  
Also note that our model predicts that \tabby~could in principle experience a comparatively rapid {\it brightening}, should an additional planet-impact event occur. 


As a final point, we note that planet-star impacts should give rise to luminous optical and X-ray transients (e.g., \citealt{Bear+11,Metzger+12}).  For Jupiter or higher mass planets, these transients may approach luminosities comparable to classical novae, placing stringent constraints on the Galactic rate of such events; however, for lower mass planet-star interactions the luminosities would be substantially dimmer and such events could more easily be missed.  

\section*{Acknowledgements} 

We acknowledge insightful conversations with Eugene Chiang and David Kipping and helpful comments on the manuscript by Smadar Naoz, Jason Wright, Lubos Neslusan, Jano Budaj, and Josh Simon.  We also thank the anonymous reviewer for helpful comments.  BDM gratefully acknowledges support from the National Science Foundation (AST-1410950, AST-1615084), NASA through the Astrophysics Theory Program (NNX16AB30G) and the Fermi Guest Investigator Program (NNX15AU77G, NNX16AR73G), the Research Corporation for Science Advancement Scialog Program (RCSA 23810), and the Alfred P.~Sloan Foundation.  KJS is supported by NASA through the Astrophysics Theory Program (NNX15AB16G)


\end{document}